\documentclass[a4paper,10pt,reqno]{amsart}
\usepackage{latexsym}
\usepackage{amssymb}
\usepackage{amsmath}
\usepackage{amsmath,amssymb,amsthm}
\usepackage{mathrsfs}
\begin{document}  

\renewcommand{\baselinestretch}{1.2}
\newcommand{\x}{\times}
\newcommand{\<}{\langle}
\renewcommand{\>}{\rangle}
\renewcommand{\i}{\infty}
\newcommand{\p}{\partial}
\newcommand{\mat}[4]{\ensuremath{{#1 \, #2 \choose #3 \, #4}}}
\newcommand{\dmat}{\left( \begin{array}{ccc}}
\newcommand{\ea}{\end{array} \right)}
\newcommand{\mbf}{\mathbf}
\newcommand{\mbb}{\mathbb}
\newcommand{\dia}{\diamond}
\newcommand{\tbf}{\textbf}
\newcommand{\rta}{\rightarrow}
\newcommand{\Rta}{\Rightarrow}
\newcommand{\mcl}{\mathcal}
\newcommand{\mclh}{\mathcal{H}}
\newcommand{\mcloi}{\mathcal{O}(-\infty)}
\newcommand{\mcls}{\mathcal{S}}
\newcommand{\mcld}{\mathcal{D}}
\newcommand{\ox}{\otimes}
\newcommand{\dg}{\dagger}
\newcommand{\wdg}{\wedge}
\newcommand{\halb}{\frac{1}{2}}
\newcommand{\prl}{\parallel}
\newcommand{\vc}[2]{(^{#1}_{#2})}
\newcommand{\inv}[1]{\frac{1}{#1}}
\newcommand{\lo}{\leqno}
\newcommand{\ex}{\begin{equation}}
\newcommand{\ey}{\end{equation}}
\newcommand{\eu}{\begin{eqnarray}}
\newcommand{\ev}{\end{eqnarray}}
\newcommand{\XXX}{$\clubsuit \clubsuit \clubsuit$}
\newcommand{\sqre}{\fbox{\rule{0cm}{.15cm}   }}
\newcommand{\grad}{\mbox{ grad }}
\newcommand{\curl}{\mbox{ curl }}

\renewcommand{\a}{\alpha}
\renewcommand{\b}{\beta}
\renewcommand{\d}{\delta}
\newcommand{\D}{\Delta}
\newcommand{\e}{\varepsilon}
\newcommand{\g}{\gamma}
\newcommand{\G}{\Gamma}
\newcommand{\io}{\iota}
\renewcommand{\l}{\lambda}
\renewcommand{\L}{\Lambda}
\newcommand{\vi}{\varphi}
\newcommand{\s}{\sigma}
\newcommand{\Sig}{\Sigma}
\renewcommand{\t}{\tau}
\renewcommand{\th}{\theta}
\newcommand{\Th}{\Theta}
\renewcommand{\o}{\omega}
\renewcommand{\O}{\Omega}
\renewcommand{\u}{\upsilon}
\newcommand{\U}{\Upsilon}
\newcommand{\z}{\zeta}
\newcommand{\kp}{\kappa}

\newcommand{\lpd}{(2\pi)^{-1/2}}
\newcommand{\kpd}{(2\pi)^{-3/2}}
\newcommand{\kpdz}{\inv{8\pi^3}}
\newcommand{\Opsc}{Op\psi c}
\newcommand{\opsc}{Op\psi c}
\newcommand{\pdo}{$\psi$do}
\newcommand{\trace}{\mbox{trace }}

\newtheorem{theorem}{Theorem}[section]
\newtheorem{lemma}[theorem]{Lemma}
\newtheorem{corollary}[theorem]{Corollary}
\newtheorem{proposition}[theorem]{Proposition}
\newtheorem{definition}[theorem]{Definition}
\newtheorem{remark}[theorem]{Remark}
\newtheorem{thm}[theorem]{Theorem}
\newtheorem{prop}[theorem]{Proposition}
\newtheorem{cor}[theorem]{Corollary}
\newtheorem{rem}[theorem]{Remark}
\newtheorem{defn}[theorem]{Definition}
\newtheorem{note}[theorem]{Note}
\newtheorem{obs}[theorem]{Observation}
\newtheorem{conj}[theorem]{Conjecture}

\title[Dirac equation of the Compton effect]
{The Dirac Equation of the Compton Effect}
\author[H.O.CORDES]{H.O.CORDES\copyright 2017}

        \bigskip

         \scriptsize

\begin{abstract}

We are looking at a Dirac electron in the electro-magnetic field of a plane 
monochrome polarized X-ray. It will be attempted to link the terms of a 
certain (joint) asymptotic expansion of the Heisenberg propagations of 
momentum- and total-energy- observables with collision of the electron with 
0, 1, 2, 3, ... photons. Our asymptotic expansion neglects terms small for 
states of very high frequencies, and might be natural, in many respects.
A special focus will be given to the single collision term. We attempt a 
description of this term as a (distribution-) integral operator, to analyze 
directionally different oscillations of the electron, to be compared with 
those of the Compton model. We obtain agreement of scattering frequencies, 
but not of scattering directions. It might be recalled, in that respect, 
that the Dirac particle is not a simple mass point --- it has a dual nature 
as electron-positron; it has a spin who possible could rotate. after the 
collision, then absorbing energy-momentum, etc.

\end{abstract}

\maketitle

\normalsize


\LARGE
\section{Introduction}

\normalsize

In this paper we shall investigate Dirac's system of partial differential equations, 
describing wave mechanics of the electron-positron. We specify the electro-magnetic field to be the (time-dependent)
field of a plane polarized X-ray wave --- so, we should have the Dirac-waves of a particle --- electron or positron ---
when irradiated by such an electro-magnetic wave.
 
This, of course, is the setup of the well known \emph{Compton experiment} (cf. [Cp1],[Cp2]).
Compton shows a frequency drop of the scattered X-ray-wave, depending on the scattering angle $\th$, and a very natural
explanation of his result arises, if the Dirac particle is thought to be hit by a 'light particle' (called \emph{photon}),
with a completely elastic collision, preserving energy and momentum. The light particle must have energy $h\nu$ and 
momentum $\frac{h\nu}{c}$, with frequency $\nu$ of the X-ray-wave, and speed of light $c$, Planck constant $h$.
These facts then are regarded as \emph{proof of the dual nature of radiation --- as a stream of photons, or else as a wave},
appearing as a contradiction.

While Compton then wants to regard both tbe X-ray-light, and the electron as particles, we, on the other hand,
want to regard both, the X-ray, and the electron, as a wave each, then looking for a different, more complicated, but
perhaps also more realistic explanation --- hopefully with no contradiction.

Just as Compton, we focus on  energy and momentum of the Dirac particle as of \emph{observable quantities}. For us they are
called $H(t)$ and $D_1\ ,\ D_2\ ,\ D_3$ , setting $D_j=-i\frac{\p}{\p x_j}$\ . We work with Heisenberg's representation,
keeping states constant and letting observables wander, according to $A\rta A_t=U^{-1}(t)AU(t)$ with the unitary 
\emph{propagator} $U(t)$ of Dirac's equation.

Our approach then is, to examine the observables $(H(t))_t\ ,\ (D_j)_t$ while trying to drop terms being negligible in size
for very high frequencies. To measure this 'negligible status' we use the well known 
chain $\{\mclh_s\ :\ -\i\leq s\leq\i\}$ of $L^2$-Sobolev spaces, introducing a class of operators continuous from 
$\mclh_s\rta\mclh_{s-m}$ for all $s$ , where $m$ is a given real. Such operators are said to be of (differentiation) order $m$.

Notice that, in a sum $A+B$ of two operators of order $m$ and $m'$, resp., where $m>m'$, the 
operator $B$ indeed might become negligible, if it comes to high frequencies.

In the momentum representation --- that is, after applying the (unitary) Fourier transform
$u(x)\rta u^\wdg(\xi)=(Fu)(\xi)$, the Sobolev norms may be written as
$$
\|u\|_s=\|\<\xi\>^s u^\wdg(\xi)\|_0\ ,\ \mbox{  with    } \<\xi\>=\sqrt{1+|\xi|^2}\ ,\ 
\lo{(1.1)}
$$
It then follows that for the Fourier transformed sum $A^\wdg+B^\wdg$ we may write $B^\wdg=C^\wdg \<\xi\>^{-r}$ , where $r=m-m'>0$,
and with an operator $C$ of order $m$. A high frequency state $\psi_0$ should have a $\psi_0^\wdg(\xi)$
 with 'action at large $|\xi|$ only '
 --- its $\psi_0^\wdg(\xi)$ should vanish in a large ball $|\xi|\leq R$, so, that, in the sum 
$(A+B)^\wdg=A^\wdg+C^\wdg\<\xi\>^{-r}$ the contribution of $B$ should be negligible.

Using this interpretation, it turns out that the observables $(H(t))_t\ ,\ (D_j)_t$ will have an ' asymptotic expansion
modulo $\mclh_{-\i}$ ' . 

These operators all are of order $1$, we may formally write 
$$
A=A_0+A_1+A_2+\cdots \mbox{  (mod } \mclh_{-\i} )\ ,\ \mbox{  with  } A_j \mbox{  of order  } 1-j\ ,
\lo{(1.2)}
$$
for each of $A=(H(t))_t\ ,\ A=D_j$, meaning that 
$$
A-\sum_{l=0}^N A_l \mbox{  is of order  } -N  \mbox{  for all  } N=0,1,\cdots \ ,\ 
\lo{(1.3)}
$$
while no convergence of the infinite sum is implied.

\bigskip

Clearly, however, such expansion establishes an ordering of the operator $A$ regarding its significance for high frequencies:

If the sum, at right, is cut off at the N-th term, then all following terms are negligible, looking at states of sufficiently high
frequencies.

We then propose to look at the zero-th term to represent the case of 'no collision with a photon' , with the first term
representing 'collision with one photon', etc.  

We let our X-ray propagate in the positive $x_1$-direction. Then we get
$$
(D_j)_t=D_j+A^0_{jt}+\cdots \mbox{  (mod }\mclh_{-\i})\ ,\ 
(H(t)_t=H(t)+A^0_{1t}+\cdots \mbox{  (mod }\mclh_{-\i})\ ,\ 
\lo{(1.4)}
$$
where $A^0_{2t}=A^0_{3t}=0$.

The case of a single photon collision then should be represented by the operator $A^0_{1t}$ occurring in both expansions,
of $(H(t))_t$ and $(D_1)_t$.

This operator then will be investigated more closely.

Note, we are approaching this problem from the side of the Dirac wave --- while, of course, Compton looks at the X-ray wave.
Correspondingly, we should look at Comptons electron-photon collision from the side of the electron, as is done in sec.5.
We also will obtain information on the (distribution-) integral kernel of this $A^0_{1t}$, in order to study its oscillatory
behaviour in configuration space.

We find an interesting coincidence of (directionally dependent) oscillation frequenies. But in other respects the comparison
does not seem translucent --- we might have to think of the fact that the Dirac particle is not just a mass point: It has
a mechanical moment and a magnetic moment (investigated in the early chapters of [Co7]). There also is the split of
electron and positron states, possibly influencing this.

In this latter respect (and our previous investigations) the observables $H(t)$ and $D_j$ are not \emph{precisely predictable},
and we may find pp-corrections hidden in the expansion (1.2) --- though not in the term $A^0_{1t}$.

The derivation of expansions (1.4) has been layed out in sec.'s 8,9,10, although this was discussed under similar aspects in [Co7].
The proof of thm.6.1 --- at the foot of our integral operator representation of $A^0_{1t}$ amounts to a lengthy calculation, 
involving formulas for Bessel functions available best in the collections [MOS]. These details, and a variety of other estimates
are skipped, although we might feel obligated to offer details in a follow-up publication.

\LARGE

\section{Our Setup}

\normalsize 

We are using Dirac's equation in its \emph{non-relativistic form}, writing\footnotemark
$$
\frac{\p\psi}{\p t} +iH\psi=0\ \ \ ,\ \ \ H=\sum_{j=1}^3\a_j(D_j-\mcl{A}_j)+\b +\mcl{V}\ \ \ ,\ \ \ D_j=\inv{i}\frac{\p}{\p_{x_j}}\ ,
\lo{(2.1)}
$$
\footnotetext{The physical constants
usually found in the Dirac equation have been absorbed by choosing
proper units: The unit of length is the \emph{Compton wave length} of
the electron $\hbar/mc \approx 3.861 \x 10^{-13} m$. The unit of time is
$\hbar/mc^2 \approx 1.287 \x 10^{-21} sec $. The unit of energy is
$mc^2 \approx 0.5 MeV$. This will make $c=m=\hbar=|e|=1$. Furthermore, we must
choose units of electromagnetic field strength to absorb the factor $e$
- rather $|e|$ - the elementary charge (while $e$ (of course) counts as a
negative charge). Note that, with these units,
we get $\mcl{E}=-$grad $\mbf{V}-\mbf{A}_{|t}\ ,\ \mcl{B}=$curl $\mbf{A}$
as electrostatic and magnetic
field strength, resp. Also, for the Coulomb potential we get
$\mbf{V}(x)=-\frac{c_f}{|x|}$ with the fine structure constant
$c_f\approx \inv{137}$\ .}
with certain $4\x 4$-Dirac-matrices\footnotemark $\a_j,\b$ and the special electro-magnetic potentials 
$$
\mcl{V}=0\ ,\ \mcl{A}=(\mcl{A}_1,\mcl{A}_2,\mcl{A}_3)\ ,\ \mcl{A}_1=\mcl{A}_3=0\ ,\ \mcl{A}_2=\e_0\sin\o(x_1-t)\ .
\lo{(2.2)}
$$

\footnotetext{Our Dirac matrices $\a_j,\b$ are self-adjoint $4\x 4$-matrices satisfying  
$\a_j\a_l+\a_l\a_j=2\d_{jl}\ ,\ \a_j\b+\b\a_j=0\ ,\ j,l=1,2,3.$}

The above (2.1) represents a first order symmetric hyperbolic system of 
4 partial differential equations for the 4 unknown complex-valued functions
$\psi(t,x)=(\psi_1,\psi_2,\psi_3,\psi_4)(t,x)$ in the 4 real variables $t,x=(x_1,x_2,x_3)$.

The electro-static potential $\mcl{V}$ vanishes identically, the electro-magnetic potential $\mcl{A}$ will correspond to the
field
$$
\mcl{E}=\o\e_0\cos\o(x_1-t) (0,1,0)\ ,\ \mcl{B}=\o\e_0\sin\o(x_1-t)(0,0,1)\ 
\lo{(2.3)}
$$
of a plane electro-magnetic wave (of frequency $\nu=\o/2\pi$) propagating in the $x_1$-direction, 
with electric and magnetic fields oscillating in the $(x_1,x_2)$-plane and
$(x_1,x_3)$-plane, respectively. (Recall, we have $\mcl{E}=-\grad\mcl{V}-\p\mcl{A}/\p t\ =-\p\mcl{A}/\p t\ ,\ \mcl{B}=\curl\mcl{A}$.)

In the sense of (old-fashioned) Schroedinger-type wave-mechanics, the underlying Physics of this problem should be that of a
single electron (or positron) propagating in the field of that electro-magnetic wave --- an X-ray wave, if the circular frequency
$\o$ is properly chosen.

    This, of course, is the setup of the Compton effect --- essential in proving the dual particle-wave-property of the radiation,
in effect, the existence of Photons. We are guided by the conjecture that this dual nature of light is nothing very mysterious,
rather that it simply comes out of theory of partial differential equations, that --- for very large frequencies--- this
encounter between light and Dirac particle, just has this property of discrete collision between particles.

For the Physics of this problem we first must solve the Dirac equation's \emph{initial-value problem:} 
For a given function $\psi_0(x)$
there exists a unique function $\psi(t,x)$ with $\psi(0,x)=\psi_0(x)$ solving our Dirac equation 
$\frac{\p\psi}{\p t} +iH(t)\psi=0$ . The assignment $U(t):\psi_0(x)\rta \psi(t,x)$ then defines a linear operator U(t),
called the \emph{propagator} of our problem.

Introducing the Hilbert space $\mcl{H}$ of squared integrable functions with norm $\|\psi_0\|=\{\int dx|\psi_0|^2dx\}^\halb$
the propagator $U(t)$ will be a unitary operator. The functions $\psi_0\in \mcl{H}$ with norm 1 
will define the \emph{ (physical) states }
of the Dirac particle --- electron or positron. Observable quantities --- shortly called \emph{ 'observables' } will be 
represented by unbounded self-adjoint linear operators of $\mcl{H}$.
 Specifically, location and momentum, the two quantities guiding the
classical propagation of a mass point are represented (respectively) by 
$$
\mbox{  multplication }\psi_0(x)\rta x_j\psi_0(x)=M_j\psi_0\ ,\ 
\mbox{    and differentiations    }\psi_0(x)\rta\inv{i}\frac{\p\psi_0}{\p x_j}(x)=D_j\psi_0\ ,\ j=1,2,3 \ .
\lo{(2.4)}
$$

\begin{center}

\bigskip

For an observable $A$ and a state $\psi_0$ one then will be able

 to predict \emph{a statistical expectation value} $\breve{A}_t$, at time $t$, setting

\bigskip

\end{center}

$$
\breve{A}_t= \<\psi_t,A\psi_t\> \mbox{  with the state } \psi_t(x)=\psi(t,x) 
\mbox{  and the inner product  } \<.,.\> \mbox{  of } \mclh.
\lo{(2.5)}
$$
Or, using our unitary propagator $U(t)$, noting that $\psi_t=U(t)\psi_0$, we get 
$\<\psi_t,A\psi_t\>=\<U(t)\psi_0,AU(t)\psi_0\>=\<\psi_0,U^{-1}(t)AU(t)\psi_0\>$. Setting $A_t=U^{-1}(t)AU(t)$ we may write (2.5) as
$$
\breve{A}_t= \<\psi_0,A_t\psi_0\>\ .
\lo{(2.5')}
$$

\bigskip

Note, the above scheme was designed for Schroedingers equation --- or at least for wave equtions describing a single particle of 
a given kind. But the Dirac equation is a wave equation for two different kinds of particles --- electrons and positrons.
Unless we are careful in selecting our observables we shall be thrown into some bad contradictions. Generally, a state $\psi_0$
will be a mix of electron states and positron states. A self-adjoint operator $A$ is allowed to represent an observable only
if it does not mix up electron states and positron states --- in a sense to be specified. We have analyzed these things more
carefully in [C01],[Co5],[Co6],[Co7], introducing a kind of observables we call \emph{precisely predictable} (abbrev. pp-observables).

In the present paper we shall be focusing on two special observables --- the total energy $H(t)$ and (the components$D_j$ of)
the mechanical momentum. Both of these are not precisely predictable but will become 'pp' if a small correction is added. This is
to be kept in mind in the following.

\LARGE

\section{Asymptotic expansions modulo $\mclh_{-\i}$}

\normalsize

Here we recall the $L^2$-Sobolev spaces\footnotemark $\{\mclh_s\ :\ -\i<s<\i\}$ : For  a 
nonnnegative integer $s=k$, the space $\mclh_k$ consists of all
functions in $\mclh=\mclh_0=L^2(\mbb{R}^3)$ having all 
derivatives of order $\leq k$ in $L^2$\ .

\footnotetext{It is practical here to deal with temperate distributions $u(x)$ instead of squared integrable functions, just
to avoid having to deal with strong $L^2$-derivatives, etc. Accordingly, our derivatives 
here should be regarded as distribution derivatives,
and the Fourier transform is a transform of the space of temperate distributions.}

The spaces can be made Hilbert spaces, and their definition can be interpolated and extended to all real $s$ if we 
recall the Fourier transform
$$
Fu(\xi)=u^\wdg(\xi)=\kpd\int dx e^{-ix\xi}u(x)\ ,\ F^{-1}u(x)=\bar{F}u(x)=u^\vee(x)\ ,\ 
\lo{(3.1)}
$$
as a unitary operator of $\mclh$ diagonalizing the components $D_j$ of the momentum : we have 
$$
FD_jF^{-1}u(\xi)=\xi_ju(\xi) \mbox{  = multiplication by } \xi_j\ .
\lo{(3.2)}
$$
For a function $f(\xi)$ we then introduce the operator $f(D)u(x)=(F^{-1}f(\xi)Fu)(x)$. Then with the function 
$\<\xi\>=\sqrt{1+\xi^2}$ we find that $\mclh_k=\{u\ :\ \<D\>^ku\in\mclh\}$. Generalizing, we then introduce the Hilbert norm
and inner product
$$
\|u\|_s=\|\<D\>^su\|\ \ \ \ \ ,\ \ \ \ \ \<u,v\>_s=\<\<D\>^su,\<D\>^sv\>
\lo{(3.3)}
$$
with norm $\|.\|$ and inner product $\<.,.\>$ of $\mclh=\mclh_0=L^2$.

This defines a decreasing chain of spaces $\mclh_s$, as $-\i<s<\i$, we extend it by adding $\mclh_\i=\cap \mclh_s\ ,\ 
\mclh_{-\i}=\cup\mclh_s$. We then get
$$
\mclh_\i\ \subset\ \mclh_s\ \subset\ \mclh_t\ \subset\ \mclh_{=\i}\ ,\ \mbox{  as  } s>t\ .
\lo{(3.3')}
$$

A continuous linear operator $A: \mclh_s\ \rta\ \mclh_t$ is well defined by its restriction to $\mclh_\i$  , then interpreted
as a map $\mclh_\i\ \rta\ \mclh_{-\i}$, since $\mclh_\i$ is dense in each $\mclh_s$ of finite $s$.

We then introduce an \emph{order} for certain linear operators $A:\mclh_\i\rta\mclh_{-\i}$. Such an operator $A$ is said to be
of order $m$ if it induces continuous maps $\mclh_{s-m}\ \rta\ \mclh_s$ for every $s\in\mbb{R}$. Here the order $m$ may be an
arbitrary real.

The differentiations $D^\th=D_1^{\th_1}D_2^{\th_2}D_3^{\th_3}$ are examples of operators of order $|\th|=\th_1+\th_2+\th_3$, and
the order $m$ may frequenctly be referred to as \emph{differentiation order}.

Note, such operators also me be regarded as (unbounded closed) operators $A:\mclh_m\cap\mclh\ \rta\ \mclh$ 
having $\<D\>^sA\<D\>^{m-s}$ bounded, implying that also $\<D\>^{m-s}A^*\<D\>^{s}$ be bounded, for all $s$ or,
$\<D\>^{t}A^*\<D\>^{m-t}$ bounded for all $t$, with $A^*$ denoting the $L^2$-adjoint of $A$.

Thus, if an operator $A:\mclh_\i\ \rta\ \mclh_{-\i}$ is of differentiation order $m$ then also its $L^2$-adjoint is of order $m$.

\begin{defn}

An operator $A$ of order $m_0$ is said to have an asymptotic expansion (modulo $\mclh_{-\i}$), written as
$$
  A=A_0+A_1+A_2+\cdots +A_n + \ \cdots\ \ \mbox{   (mod }     \mclh_{-\i})\ ,\ 
\lo{(3.4)}
$$
where the operators $A_j$ are of order $m_j$ with $m_0 > m_1>m_2>\cdots> m_n > m_{n+1}>\ldots\ \rta -\i$, and such that
   $A-\sum_{j=0}^NA_j$ is of order $m_{N+1}$ for all $N=0,1,2,\cdots$.

\end{defn}

\bigskip

\begin{prop}

If an operator A has the asymptotic expansion (3.4) , then also its $L^2$-adjoint $A^*$ has the asymptotic expansion
$$
  A=A_0^*+A_1^*+A_2^*+\cdots +A_n^* + \ \cdots\ \ \mbox{   (mod }     \mclh_{-\i})\ ,\ 
\lo{(3.4*)}
$$

\end{prop}

\begin{obs}

Note, an asymptotic expansion modulo $\mclh_{-\i}$ does not imply any kind of convergent infinite series. But it suggest an ordering
regarding the influence of terms applied to high frequency states (or states with large $|\xi|$ considered in the 
momentum representation) --- that is, applied to wave functions $\psi_0(x)$, with Fourier transform $\psi^\wdg(\xi)$ 
vanishing in a large ball $|\xi|\leq T$\ : We get
$$
A\psi\approx A_0\psi+\cdots+A_N\psi
\lo(3.5)
$$
with accuracy of the $"\approx"$ depending on the 'largeness' of $T$.

\end{obs}

\LARGE

\section{Heisenberg Transform of Momentum and Energy}

\normalsize

Regarding the wavemechanical Physics, we now will work with (2.5'), not with (2.5), where the operator family $A_t=U^{-1}(t)AU(t)$
will be called the Heisenberg transform of the observable $A$.

As a first simplification, in that direction, note that a conjugation with the translation operator
$$
T_t=e^{itD_1}\mbox{  given as  } T_t:\psi(x)\rta\psi(x_1+t,x_2,x_3)
\lo{(4.1)}
$$
will give a time-independent operator $T_t^{-1}H(t)T_t=H(0)$. As a consequence the Dirac equation (2.1) then reduces to
$$
\frac{\p\chi}{\p t} +iK\chi=0\ \ \ ,\ \ \ K=H(0)-D_1\ \ \ ,\ \ \ \chi=T_{-t}\psi\ .
\lo{(4.2)}
$$

Consequently we get\footnotemark
$$
U(t)=T_{-t}e^{-iKt}\ ,\ \mbox{  with  } K=H(0)-D_1\ ,\ 
\lo{(4.3)}
$$
where we are able to analyze explicitly the exponential group $e^{-iKt}$ of the self-adjoint operator $K$ independent of $t$,
while the translation $T_t$ is more or less trivial.

\footnotetext{More generally, the propagator $U(\t.t)$ mapping from $\t$ to $t$ has the form $U(\t.t)=T_{-t}e^{-iK(t-\t)}T_\t$,
as easily checked.}

As a consequence of (4.3), if we write $A_t=U^{-1}(t)AU(t)$, for a general operator $A$ , we get
$$
(H(t))_t-H(0)=(D_1)_t-D_1\ .
\lo{(4.4)}
$$
Indeed, we have 
$(H(t))_t=U^{-1}(t)H(t)U(t)=e^{-iKt}T_tH(t)T_{-t}e^{iKt}$

$=e^{-iKt}H(0)e^{iKt}
=e^{-iKt}Ke^{iKt}+e^{-iKt}D_1e^{iKt}=K+(D_1)_t=H(0)-D_1+(D_1)_t\ .$

Accordingly, if we control the transforms $(D_j)_t$ of the momentum components, we also get $(H(t))_t=H(t)+(D_1)_t-D_1 $.

\begin{thm}

We have asymptotic expansions
$$
(D_j)_t=D_j+A^0_{jt}+A^1_{jt}+A^2_{jt}+\cdots \mbox{  (mod }\mclh_{-\i})\ ,\ 
(H(t)_t=H(t)+A^0_{1t}+A^1_{1t}+A^2_{1t}+\cdots \mbox{  (mod }\mclh_{-\i})\ ,\ 
\lo(4.5)
$$   
where the operators $A^l_{jt}$ are of order $-l$.

In particular, we have $A^0_{2t}=A^0_{3t}=0$, and (with $s_j(\xi)=\xi_j/\<\xi\>\ ,\ j=1,2,3, $ )
$$
A^0_{1t}=\o\e_0\{\frac{h_0(D)}{\<D\>}\int_0^t d\t \cos(\o(x_1-\t)\cos(\o\t s_1(D))
-\int_0^t d\t \sin(\o(x_1-\t)\sin(\o\t s_1(D))\}s_2(D)\ .
\lo{(4.6)}
$$
For $l=2,3,\cdots$ we have 
$$
A^l_{jt}=\sum_{k=-l-1}^{l+1}e^{i\o k x_1}a^l_{jkt}(D)\ ,\ 
\lo{(4.7)}
$$
with $a^l_{jkt}(\xi)$ of polynomial\footnotemark order $-l$ .

For l=1 we have  
$$
A^1_{jt}=\sum_{k=-2}^{2}e^{i\o k x_1}a^1_{jkt}(D) - U^{-1}(t)B_{jt}U(t)\ ,\ a^1_{jkt} \mbox{  of pol. order } -1\ ,\ 
\lo{(4.8)}
$$
with $B_jt$ (of order $-1$) representing a pp-correction of (resp. $D_j$ or $H(t)$) --- it makes those operators
'precisely predictable' . (This $B_{jt}$ has a representation similar to (4.6) again --- cf. sec.'s 9,10 for details.)

\end{thm}

\footnotetext{A function $f(\xi)$ is said to be of polynomial order $m$ if all derivatives of order $\leq k$ are $O(\<\xi\>^{m-k})$,
for all $k=0,1,\cdots$.}

The proof of thm.4.1 is discussed in sec.'s 8,9,10.

Ignoring the pp-correction term, we then want to think of $A^l_{jt}$ as of the term representing the $l+1$-fold 
collision of our Dirac particle with a photon. Note, the Fourier expansion 4.7 means that --- in the momentum representation
(i.e., after transforming with the unitary Fourier transform) the factor $e^{il\o x_1}$ goes into the momentum shift by 
$\pm l$ multiples of $h\nu/c$, illuminating the physical situation.

We, of course, will be focusing onto the case of a single such collision --- represented by the operator of (4.6).
We clearly have
$$
(D_j)_t-D_j-A^0_{jt}\ ,\ (H(t))_t-H(t)-A^0_{1t} \mbox{    of order  } -1\ .\ 
\lo{(4.9)}
$$
So, to study the single collision, we will focus on the operator (4.6).

\LARGE

\section{The Compton Collision}

\normalsize

Let us recall the electron-photon collision of the Compton effect:

Using that $\hbar=c=m_e=|e|=1$ and $h=2\pi\hbar=2\pi$ , in our units,
conservation of energy and momentum gives (cf. fig.1)

$\halb m_e v^2=2\pi \D\nu\ ,\  \mbox{  with  } \D\nu=\nu_0-\nu_1$ and
$0 + 2\pi\nu_0=v_h + 2\pi\nu_1\cos\th_r\ ,\ 0=v_v+2\pi\nu_1\sin\th_r$
for the velocity-vector $v=(v_h,v_v)$ with horizontal and vertical components $v_h\ ,\ v_v$.
(Before the collision the electron's momentum is zero, the momentum of the photon is $(2\pi\nu_0,0)$;
 After the collision the electrons momentum is $(m_ev_h,m_ev_v)=(v_h,v_v)$\ ,\ the momentum 
of the photon is $(2\pi \nu_1 \cos\th_r,2\pi\nu_1\sin\th_r)$ .) So, 
$$
v_h=2\pi(\nu_0-\nu_1\cos\th_r)\ ,\ v_v=-2\pi\nu_1\sin\th_r\ .
\lo{(5.1)}
$$

\begin{center}
\begin{picture}(200,150)(0,-40)
\put(00,50){\vector(1,0){100}}
\put(100,50){\vector(2,1){40}}
\put(100,50){\vector(1,-2){30}}
\multiput(100,50)(10,0){10}{\line(1,0){5}}
\qbezier(130,65)(136,55)(135,50)
\qbezier(140,50)(145,20)(120,10)
\put(30,40){\scriptsize{incoming X-ray}}
\put(130,75){\scriptsize{electron}}
\put(110,-20){\scriptsize{scattered X-ray}}
\put(140,20){\scriptsize{$\th_r$}}
\put(135,58){\scriptsize{$\th_e$}}
\end{picture}

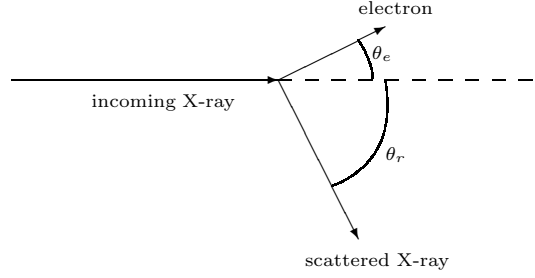
 Fig.1. Mechanical Electron-Photon Scattering
\end{center}

\bigskip

We get 
$
4\pi\D\nu=v^2=v_h^2+v_v^2=4\pi^2\{(\nu_0-\nu_1\cos\th_r)^2+\nu_1^2\sin^2\th_r\}\ ,\ 
$
that is,
$$
\D\nu=\pi\{\nu_0^2+\nu_1^2-2\nu_0\nu_1\cos\th_r\}=\pi\{(\D\nu)^2+2\nu_0\nu_1(1-\cos\th_r)\}\ .
\lo{(5.2)}
$$

Looking at the  wavelengthes $\l_0\ ,\ \l_1$ : Since we have $c=1$, we get $\l_j=\frac{c}{\nu_j}=\inv{\nu_j}$. So,
$
\D\l=\l_1-\l_0=\frac{\D\nu}{\nu_0\nu_1}\ .
$
Dividing (5.2) by $\nu_0\nu_1$ we get
$
\D\l=\pi\frac{(\D\nu)^2}{\nu_0\nu_1}+4\pi\sin^2\th_r/2\ .
$
Neglecting the first term at right we get the well known Compton formula
$$
\D\l=2\L \sin^2\th_r/2\ ,\ \L=\frac{h}{m_e c}=2\pi=\mbox{  Compton wavelength  }\ .
\lo{(5.3)}
$$

For us, the electron scattering angle $\th_e$ will be of interest. For a first approximation we get
$$
\tan\th_e=\frac{v_v}{v_h}=-\frac{\nu_1\sin\th_r}{\nu_0-\nu_1\cos\th_r}\approx-\frac{\sin\th_r}{1-\cos\th_r}\ ,\ 
=\cot(-\th_r/2)=\tan(\frac{-\pi}{2}+\th_r/2)\ ,\ 
\lo{(5.4)}
$$
assuming that $\D\nu<<\nu_0$ . This implies that 
$$
\th_e\approx -\frac{\pi}{2}+\frac{\th_r}{2}\ .
\lo{(5.5)}
$$

\ ,  Indeed (5.3) implies $\D\l\leq 4\pi$, so that
$\D\nu\leq 4\pi\nu_0\nu_1=4\pi\nu_0^2-4\pi\nu_0\D\nu$, i.e.,
$
\D\nu\leq\frac{4\pi\nu_0^2}{1+4\pi\nu_0}\ .
$
For Compton's experiment we believe we might set an X-ray-energy $\approx 13000$ e-Volt. With $m_e\approx 511000$ eVolt 
we get $\nu_0\approx \frac{1}{40}$, giving us an idea of accuracy\footnotemark of (5.4).

\footnotetext{A more accurate estimate of the relation between $\th_r$ and $\th_e$ might be given by the relation
$$
\cot\th_e=1.025\cot(-\frac{\pi}{2}+\frac{\th_r}{2})\ ,\ 
$$}

\bigskip

When now we start a wave-mechanical investigation of this problem, then we might be able to 
expose some properties of the electron-wave --- while, of course, the Compton effect looks only at the 
X-ray wave involved. 

Our above discussion of the mechanical problem entirely rests on reflections on energy and mechanical momentum
of the electron. Both of these will be determined by the velocity $v$.
From our above discussion around Fig.1 we get f'la (5.1) in the form
$$
v^2=4\pi^2\{\nu_0^2+\nu_1^2-2\nu_0\nu_1\cos\th_r\}\ ,\ 
v_h=2\pi(\nu_0-\nu_1\cos\th_r)\ ,\ v_v=-2\pi\nu_1\sin\th_r\ .
\lo{(5.1')}
$$

Just as above, we shall focus --- wave-mechanically --- on the 4 observables 
$H(t)\ ,\ D_1\ ,\ D_2\ .\ D_3$, representing energy and
momentum of the Dirac particle.

In the setup of Fig.1 we shall have to interpret $v_h$ as the $x_1$-velocity component, while $v_v$ should be 
within the $(x_2,x_3)$-plane, since $x_1$ is the direction of our X-ray.

Looking at (5.1') again, thinking of the norm $m_e|v|$ of the momentum, we might write 
$$
m_e|v|=|v|=2\pi\sqrt{\nu_0^2+\nu_1^2-2\nu_0\nu_1\cos\th_r}=h\nu^{\th_r}/c\ ,\ \mbox{  with  } 
\nu^\th=\sqrt{\nu_0^2+\nu_1^2-2\nu_0\nu_1\cos\th}
\lo{(5.1'')}
$$
looking like a frequency, dependent on $\th$ (For $\th=0$ we get $\nu^\th=\nu_0-\nu_1$\ ,\ for $\th=\pi$ we have
$\nu^\th=\nu_0+\nu_1$.) So, it seems that even the momentum of the electron may be described in the form $h\nu/c$,
with a certain frequency $\th=\th_r$. But we might think of $\nu^{\th}$ as of a frequency of an oscillation 
related to the Dirac particle's wave, 
not of some electro-magnetic wave.

Here we shall focus on the term (4.6) of thm.4.1, we believe to represent a single collision in the Heisenberg transforms of 
total energy and momentum. We shall try to get this term --- or its time-dependent part --- as an integral operator
with distribution kernel. For a particle near zero, we must assume a state vanishing outside a neighbourhood of $x=0$. 
Then, following the integral kernel, in a scattering direction $\th_e$ we will look for oscillation frequencies between $0$ and
$2\nu_0=\o/\pi$. 

Amazingly, such frequencies may be found, although we are not satisfied with their dependence on $\th_e$.

\LARGE

\section{The Integral Kernel of the Single Collision Term}

\normalsize

Looking at formula (4.6), describing the term we want to make responsible for observing a single electron-photon collision,
we shall focus on its time derivative
$$
\dot{A}^0_{1t}=\o\e_0\{\frac{h_0(D)}{\<D\>} \cos(\o(x_1-t)\cos(\o t s_1(D))
-\sin(\o(x_1-t)\sin(\o t s_1(D))\}s_2(D)\ .
\lo{(6.1)}
$$
We recall that, for an operator $f(D)$ an explicit integral operator representation is given in the form
$$
f(D)u(x)=\kpd\int dy f^\vee(x-y)u(y)\ \mbox{  with } f^\vee= F^{-1}f=\mbox{inverse Fourier transform}\ .
\lo{(6.2)}
$$
Here we will have to think of the right hand side as of a 'distribution integral' --- value of the distribution $f^\vee(x-.)$
at the testing function $u(x)$. So, we are going to focus on the inverse Fourier transform $(e^{-i\kp\xi_1/\<\xi\>})^\vee$,
to control the time-dependent parts\footnotemark of (6.1).

\footnotetext{We should be able to control the entire operator (6.1) with our technique, below, but with considerably increased
calculation efforts.}

Instead we have expliciltly calculated the transform $f^\vee(x)$ of $f(\xi)=e^{-i\kp\sin\l}$ 
with $\sin\l=\xi_1/|\xi|$, in spherical coordinates
$$
\xi_1=\rho\sin\l\ ,\ \xi_2=\rho\cos\l\cos\mu\ ,\ \xi_3=\rho\cos\l\sin\mu\ ,\ \mbox{  with  } 
0\leq \rho<\i\ ,\ 0\leq\mu <2\pi\ ,\ |\l|\leq\pi/2\ ,
\lo{(6.3)}
$$
claiming the difference between $(e^{-i\kp\xi_1/\<\xi\>})^\vee$ and $(e^{-i\kp\xi_1/|\xi|})^\vee$ unimportant --- to be
shown later on.

The transform $(e^{-i\kp\sin\l})^\vee(x)=(cos(\kp\sin\l))^\vee(x)-i(\sin(\kp\sin\l))^\vee(x)$ may not be calculated by an
ordinary Fourier integral. Instead, we calculate it as $-\D_x(\inv{\rho^2}e^{-i\kp\sin\l})^\vee(x)$ with the Laplace operator 
$\D_x=\sum\p_{x_j}^2$, and distribution derivatives to be switched to the kernel by partial integration. One finds that,
indeed, $f^\vee(x)$ is a genuine distribution --- with a delta-function character --- at the $x_1$-axis , i.e., $x_2=x_3=0$ ---
that is, in the direction of the radiation. For other $x=(x_1,x_2,x_3)$ with $(x_2,x_3)\neq 0$, the value $f^\vee(x)$
is an infinitely differentiable function, explicitly described by the theorem, below:

\begin{thm}

For $x$ with $(x_2,x_3)\neq 0$ we get
$$
(\cos(\kp\xi_1/|\xi|))^\vee(x)=-\inv{r^3\cos\th}h_1^e(\th)\ ,\ 
(\sin(\kp\xi_1/|\xi|))^\vee(x)=-i\inv{r^3\cos\th}h_1^o(\th)\ ,\
\lo{(6.4)}
$$
now working in spherical coordinates for $x$, i.e.
$$
x_1=r\sin\th\ ,\ x_2=r\cos\th\cos\vi\ ,\ x_3=r\cos\th\sin\vi\ ,\ \mbox{  with  } 
0\leq r<\i\ ,\ 0\leq\vi <2\pi\ ,\ |\th|\leq\pi/2\ ,
\lo{(6.3')}
$$
where $(x_1,x_2)\neq 0$ means $\th\neq\pm\pi/2$ --- i.e. $\cos\th\neq 0$ .

In (6.4) we have set 
$$
h^e(\kp,\th)=-Y_0(\kp\cos\th)\ ,\ h^e_1=(\cos\th h^e_{|\th})_{|\th}\ ,\ 
\lo{(6.5)}
$$
$$
h^o(\kp,\th)=-\mbox{ sgn}(\th)\{Y_0(\kp\cos\th)+\frac{2}{\pi}\int_{1/\cos\th}^\i \frac{\cos(\t\kp\cos\th)}{\sqrt{\t^2-1}}d\t\} \ ,\ \ 
h^o_1=(\cos\th h^o_{|\th})_{|\th}\ ,
\lo{(6.6)}
$$
using the Bessel function $Y_0(z)$ (cf.[MOS],p.66).

\end{thm}

Note, that both expressions $h^e$ and $h^o$ are smooth functions of $\th$, even at $\th=0$, where the factor of $\mbox{ sgn}(\th)$
vanishes at $\th=0$ and has special properties implying smoothness of $h^o$. Still it is technically advisable 
to also avoid the plane $x_1=0$ where we have $\th=0$. --- This is where the Dirac particle is only 'grazed' by the photon.

The proof of thm.6.1 follows the above-mentionned path: It amounts to a calculation of integrals, using old well known properties
of Bessel functions --- we have relied on the collection [MOS] of formulas for Mathematical Physics, in many respects.
Details may be published elsewhere, together with facts on comparison of $e^{-is_1(\xi)}$ with $e^{-i\sin\l}$ and estimates
verifying the things, below, in this section.

We will assume '  large times $t$  ', so that $\kp=\o t$ also will be large. Then, it turns out, we shall have
$$
\inv{\cos\th}h^e_1\approx -\kp^2\sin^2\th Y_0''(\kp\cos\th)\ ,\ 
\inv{\cos\th}h^o_1\approx -i\kp^2\mbox{ sgn}(\th)\sin^2\th Y_0''(\kp\cos\th)\ .\ 
\lo{(6.7)}
$$
For this estimate we have dropped all terms dominated by $\kp^2$. In particular, the term
$\int_{1/\cos\th}^\i \frac{\cos(\t\kp\cos\th)}{\sqrt{\t^2-1}}d\t$ and its $\th$-derivatives may be neglected ---
all involving us in lengthy estimates, to be discussed independently in a later paper.

Here we will apply some well known formulas on Bessel functions, getting (cf. [MOS],p.67 and p.139))
$$
Y_0''(z)=\halb(Y_2(z)-Y_0(z))
\lo{(6.8)}
$$
 and the Hankel-asymptotic estimates 
$$
Y_n(z)= \sqrt{\frac{2}{\pi z}}\{\sin(z-n\frac{\pi}{2}-\frac{\pi}{4})+O(\inv{|z|^{3/2}})\ \ \ ,\ \ \  n=0,1,\ldots\ .
\lo{(6.9)}
$$
\begin{cor}

Under our present assumptions --- for large $\kp$, and $x$ away from the $x_1$-axis and from the plane $x_1=0$ --- we get
$$
(\cos(\kp\xi_1/|\xi|))^\vee(x)\approx-\frac{\kp^{3/2}}{r^3}\sqrt{\frac{2}{\pi}}\frac{\sin^2\th}{\sqrt{\cos\th}}
\ \ \sin(\kp\cos\th-\frac{\pi}{4})\ ,\ 
\lo{(6.10)}
$$
$$
(\sin(\kp\xi_1/|\xi|))^\vee(x)\approx-i\mbox{ sgn}(\th)\frac{\kp^{3/2}}{r^3}\sqrt{\frac{2}{\pi}}\frac{\sin^2\th}{\sqrt{\cos\th}}
\ \ \sin(\kp\cos\th-\frac{\pi}{4})\ ,\ 
$$
\end{cor}

\LARGE

\section{Scattering Frequencies after a Single Collision}

\normalsize

With our present control on the integral kernel of the operator (6.1) let us 
examine our single collision, for a state located near $x=0$ ---
that is , we assume our state function $\psi_0(x)$ vanishing outside a 
sphere $|x|\leq \e$. The time-dependent components of the operator (6.1)
(approximately) have integral kernels
$$
\kpd\cos\o(x_1-t)(\cos(\o t\xi_1/|\xi|))^\vee(x-y)\ ,\ 
\kpd\sin\o(x_1-t)(\sin(\o t\xi_1/|\xi|))^\vee(x-y)\ .
\lo{(7.1)}
$$
When applying this integral to our state $\psi_0(y)$ we may assume $y\approx 0$. So, at an observation point $x$ far away from $0$
we may think of the spherical coordinates centered at $x=0$. Then, looking at the time-dependence only, we get
the products
$$
t^{3/2}\cos(\o(x_1-t))\sin(\o t\cos\th -\frac{\pi}{4})\ ,\ \mbox{  and  }
t^{3/2}\sin(\o(x_1-t))\sin(\o t\cos\th -\frac{\pi}{4})\ .
\lo{(7.2)}
$$
With well known trigonometric formulas, it then is evident, that this gives a superposition of oscillations of frequencies
$$
\nu_+=\nu_0(1+\cos\th) \mbox{  and  } \nu_-=\nu_0(1-\cos\th) \mbox{  with }\nu_0=\frac{\o}{2\pi}\ .
\lo{(7.3)}
$$
Here $\th$ denotes the angle between the ray 0---x and the plane $x_1=0$, the electron scattering angle would be $\frac{\pi}{2}-\th$.
With our Fig.1 and the relation (5.5) between radiation and electron scattering angles we would have to replace $\th$ in (7.3)
by $-\th_r/2$. We conclude:

\begin{quotation}

Our above oscillation frequencies seem to coincide with the electron frequencies of (5.1''), insofar as --- essentially --- 
they also extend between $0$ and $2\nu_0$. The dependencies on the scattering angle do not coincide, however.

In the latter respect,we must keep in mind that the Compton construction assumes an electron at zero location and zero momentum,
while (in our present 'wave-mechanics' ) we only have a location at zero 
with mixed momenta --- location closer to 0 involves momenta with larger and larger values.

Also, we must specify a pure electron state --- excluding positrons --- presently we have that not under consideration.

Actually, our frequencies of (5.1) present themselves as a vector $v=(v_h,v_v)$, reminding us of the fact, that our 
Dirac particle is much more than a mass point: we know, that it has a spin --- a mechanical moment and a magnetic moment
(cf. [Co7]). Thinking of it as an oriented little ball, we might think of a (periodic) rotation induced by our 
Photon collision --- in addition to  the induced momentum. Or, rather, there might be other physical properties of this
'object' of a nature, not comparable to objects of our macroscopic surrounding.

We may involve Comptons argument, allowing the electron at $0$ to have a momentum $\g$ . Modifying the discussion around
fig.1 , using spherical coordinates, we the same frequencies, but a different directional dependence. Also, that dependence
might be influenced by the other operators of the form $f(D)$ in formula (4.6).

In the following sections we shall discuss further details of our setup.

\end{quotation}

\LARGE

\section{We Look at $e^{-iKt}$ acting on $\mclh_s$}

\normalsize

Notice, the operator $K$ is (precisely) self-adjoint in $\mclh=\mclh_0$ --- as a sum $h_0(D)-D_1 +\e_0\a_1\sin(\o x_1)$ with
$H_0-D_1$ precisely self-adjoint in the domain $Dom\ =\ \mclh_1$ --- being diagonalized by the Fourier transform, while 
the term of multiplication by $\e_0\a_1\sin(\o(x_1)$ makes a bounded self-adjoint perturbation.
Accordingly, the operator $e^{-iKt}$ is well defined as a group of bounded (unitary) operator in $L(\mclh_0)$ ,just using the 
spectral theorem. The same can be stated about the operator $K$ in any of the spaces $H_s$. Indeed, we get 
$\|u\|_s=\|\<D\>^su\|_0$ showing the operator $\<D\>^s$ as an isometry $\mclh_s\ \leftrightarrow\ \mclh_0$. Setting 
$\<D\>^su=w\ ,\ \<D\>^sv=z$ one has
$$
\<u,Kv\>_s=\<\<D\>^su,\<D\>^sKv\>_0=\<w,\<D\>^sK\<D\>^{-s}z\>_s\ ,\ \mbox{  for  } u,v\in \mcl{S}=\mclh_{-\i}\ ,
\lo{(8.1)}
$$
so $K$ is represented by $K_s=\<D\>^sK\<D\>^{-s}$ over the Hilbertspace $\mclh$. We then get $K_s^*=\<D\>^{-s}K\<D\>^s$.

Regarding the commutator $[\<D\>^s,K]$, we must look at $[\<D\>^s,\sin(\o x_1)]$, since $\<D\>^s$ evidently commutes with
$h_0(D)$ and with $\a_2$.

\begin{prop}

For any function $c(\xi)$ we have
$$
[\sin(\o x_1),c(D)]=\frac{i}{2}\{e^{-\o x_1}(c(D+\o e^1)-c(D))+e^{-i\o x_1}(c(D)-c(D-i\o e^1))\}\ ,\ 
\lo{(8.2)}
$$
with $e^1=(1,0,0)$.

\end{prop}

The proof is a calculation (cf. also [Co7], formula (9.7) and prop.9.1 there).

We get 
$$
K_s=\<D\>^sK\<D\>^{-s}=K+\e_0\a_2\{[\<D.\>^s,\sin(\o x_1)]\<D\>^{-s}\}\ ,\ 
\lo{(8.3)}
$$
where the second term, at right, is an $L^2$-bounded operator. Indeed, using (8.2) on the commutator we get
$\<\xi+\o e^1\>^s-\<\xi\>^s=\int_0^\o d\t \p_\t(\<\xi+\t e^1\>^s)$ where
$\p_\t(\<\xi+\t e^1\>^s)=\p_\t((\<\xi\>^2+2\t\xi+\t^2)^{s/2})=s\<\xi+\t e^1\>^{s-2}(\xi_1+\t)$, showing that
$(\<\xi+\o e^1\>^s-\<\xi\>^s)\<\xi\>^{-s}=s\int_0^\o d\t\frac{\xi_1+\t}{1+\xi^2+2\t\xi_1+\t^2}$ 
is a bounded function of $\xi$' of polynomial order $0$. Similar for the other term.

So, again, the operator $K_s$ in the domain  $Dom\ =\ \mclh_1$ differs from the self-adjoint $h_0(D)-D_1$ by a 
operator bounded over $\mclh$, although this bounded perturbation no longer needs to be self-adjoint. Still, this 
implies existence of the group $e^{-iK_st}$ as a group of $L^2$-bounded operators, implying $\mclh_s$-boundedness of 
$e^{-iKt}$, as follows from the discussion in Ch.9.1 of Kato, [Ka1], p.478f.

\LARGE

\section{An Asymptotic Expansion of $A_t=e^{iKt}Ae^{-iKt}$.}

\normalsize

In this section we depart from the observation that the operator $A_t=e^{iKt}Ae^{-iKt}$
 satisfies 
$$
\frac{d}{dt} A_t=i[K,A_t]\ \ \ \ ,\ \ \ \ A_0=A\ \ \  ,\ \ \  [K,A_t]=KA_t-A_tK\ ,
\lo{(9.1)}
$$
apparently an initial-value problem for a first order ODE.

Assuming $A=a(x,D)$ and $A_t=a_t(x,D)$ to be pseudodifferential operators\footnotemark 
($\psi do$-s) in the algebra $Op\psi q$ (cf.[Co7], sec.3 )
we may translate (9.1) into this:

\footnotetext{A smooth function $f(x)$ is said to be of \emph{polynomial growth} --- of order $m$ if we have 
$\p^\th_xf(x)=O(\<x\>^{m-|\th|})$ for all derivatives $\p_x^\th$. The algebra $\psi q$ consists of all functions $a(x,\xi)$
being of polynomial growth (any order m) in $\xi$, uniformly in the variable $x$ and with all its $x$-derivatives --- the order
$m$ independent of $x$-derivative. That is, $\p_x^\io\p_\xi^\th a(x,\xi)=O(\<\xi\>^{m-|\th|})$ for all $\io,\th$ and all $x,\xi$,
with some real $m$ independent of $\io,\th$. Then a $\psi do$ $a(x,D)$ may be defined setting
$a(x,D)u(x)=\inv{(2\pi)^3}\int d\xi\int dy e^{i\xi(x-y)}a(x,\xi)u(y)$. 
The algebras $\psi q$ and $Op\psi q$ were explained in detail in [Co7],sec.3, also, refer to [Co1].}

$$
\dot{a}_t(x_1,\xi)=i[h_0(\xi),a_t(x_1,\xi)]
+(\a_1-1)a_{t|x_1}(x_1,\xi) + (Za_t)(x_1,\xi)\ \ \ ,\ \ \ a_0(x,\xi)=a(x,\xi)
\lo{(9.2)}
$$
$$
\mbox{ with      } (Zc)(x_1,\xi)=-i\e_0\sin\o x_1 [\a_2,c(x_1,\xi)]
+\frac{\e_0}{2}\a_2(Xc)(x_1,\xi)\ ,\
$$
$$
Xc(x_1,\xi)=\{(c(x_1,\xi+\o e^1)-c(x_1,\xi))e^{i\o x_1}
                     +(c(x_1,\xi)-c(x_1,\xi-\o e^1))e^{-i\o x_1}\}\ .
\lo{(9.3)}
$$
This equation (9.2) is a system of partial differential equations in the variables $t$ and $x_1$. It also is a commutator equation,
involving the commutator $[h_0,a_t]$. It also involves a 'functional' operator $X$ adressing the variables $\xi$.

One may expect the solution of (9.1) to be unique --- determining the function $A_t$ when $A$ is given. So, instead of
solving the Dirac equation for the state $\psi_0$ we aim at solving (9.1) --- that is, solving (9.2), 
making the observable vary, not the state.

We focus only on the 3 Momentum coordinates $D_1\ ,\ D_2\ ,\ D_3$. Trivially they all are operators of (differentiation) order $1$,
in the sense of sec.3.

We remind of the fact that the algebra $Op\psi q=\cup Op\psi q_m$ is a graded algebra --- involving the differentiation order ---
the growth-order in the $\xi$-variable. It is known that the operators of $Op\psi q_m$ are of order $m$ , in the sense of sec.3.
We shall attempt to solve (2) by first omitting terms of lower order, thus starting an iteration, leading precisely into an
asymptotic series of the form (3.4), with terms $A_j$ explicitly given (or calculable).

Note, if we assume $a_t$ and $\dot{a}_t=\p_t a$ of order m, then all terms in (9.2) are of that order (or less), except the
term $[h_0,a_t]$, formally being of order $m+1$. So, we might look at (9.2) as a condition for this term also to be of order $m$

To accomudate this commutator $[h_0,a_t]$ we introduce the 'projections' 
$$
p_\pm(\xi)=\halb(1\pm \frac{h_0(\xi)}{\<\xi\>})
\lo{(9.4)}
$$
of the spectral decomposition\footnotemark of $h_0(\xi)$.

\footnotetext{ For each $\xi$ the $4\x 4$-matrix $h_0(\xi)$ has eigenvalues $\pm\<\xi\>$ of multiplicity 2 each; the $p_\pm(\xi)$
are the orthogonal projections onto the eigenspaces. To analyze the commutator $[h_0(\xi),B]$, for any $4\x 4$-matrix $B$, we 
introduce 
$$
B^+=p_+Bp_+\ ,\ B^-=p_-Bp_-\ ,\ B^\pm=p_+Bp_-\ ,\ B^\mp=p_-Bp_+\ ,\ \
$$
then, for all $B$, we get
$$
B=B^++B^-+B^\pm+B^\mp\ ,\ [h_0,B^+]=[h_0,B^-]=0\ ,\ [h_0,B^\pm]=2\<\xi\>B^\pm\ ,\ [h_0,B^\mp]=-2\<\xi\>B^\mp\ ,\ 
[h_0,B^\pm]=[h_0,B^\mp]=0\ .
$$}

Writing 
$$
q_t(x,\xi)=a_t^+(x,\xi)+a_t^-(x,\xi)\ ,\ z_t(x,\xi)=a_t^\pm(x,\xi)+a_t^mp(x,\xi)\ ,\ 
\lo(9.5)
$$
in the sense of footnote 9 we conclude that
$$
a_t(x,\xi)=q_t(x,\xi)+z_t(x,\xi)\ ,\ \mbox{  where  } [h_0,q_t]=0\ ,\ [h_0,z_t]_+=0\ ,\ q_t\in\psi q_m\ ,\ z_t\in\psi q_{m-1}\ .
\lo{(9.6)}
$$
The property of $z_t\in\psi q_{m-1}$ follows if we 'left-right multiply' (9.2) by $p_+$ and $p_-$ (resp. $p_-$ and $p_+$), using that
$p_+[h_0,a_t]p_-=[h_0,a_t^\pm]=2\<\xi\>a_t^\pm$ , $p_-[h_0,a_t]p_+=[h_0,a_t^\mp]=-2\<\xi\>a_t^\mp$, showing that both $a_t^\pm$ and
$a_t^\mp$ must be in $\psi q_{m-1}$

\bigskip

Remembering that (2) is an equation for a $4\x 4$ matrix-function $a_t$
we distinguish three steps, to be iterated infinitely:

\begin{quotation}

Step I: We omit some lower order terms in (9.2), then trying to solve that as a
sharp equation.

Step II: We multiply the (simplified)  (9.2) left and right by $p_+$ (and left and right by
$p_-$) obtaining two differential equations to be solved. That will
get us an approximate $q_t$.

Step III: We multiply (9.2) left and right by $p_+$ and $p_-$, respectively
(or by $p_-$ and $p_+$, resp.). That will give us equations to obtain an
approximate $z_t$.

\end{quotation}.

These steps, applied alternately, in iteration, will result
in an infinite sequence of
improvements satisfying eq. (9.2) modulo $\psi p_{m-j}$ only, for
$j=1,2,\ldots$. Then an asymptotic limit\footnotemark (mod $\psi q_{-\i}$ in the sense of [Co7], prop.3.6)
must be taken to obtain an $a_t^\i=q_t^\i+z_t^\i$
solving (9.2) modulo $\psi p_{-\i}$.

\footnotetext{An infinite series $\sum_{j=0}^\i a_j(x,\xi)$ of symbols $a_j$ of order $m_j\searrow -\i$ is said to have
'asymptotic limit $a(x,\xi)$ ' if $a-\sum_{j=0}^Na_j$ is of order $m_{N+1}$ for all $N=0,1,\cdots$. In [Co7] we have discussed 
the fact that, \emph{for every sequence} $\{a_j\in \psi q_{m_j} : j=0,1,\cdots\}$ \emph{the series} $\sum a_j$
\emph{ has such an asymptotic limit} $a(x,\xi)$, of order $m_0$, assuming that
$m_j\searrow -\i$ .}

With such $a_t^\i(x_1,\xi)\in\psi p_m$ we then define the operator
$A_t^\i=a_t^\i(x_1,D)$, and then define
$$
B_t=e^{-iKt}A_t^\i e^{iKt}-A_0^\i\ .
\lo{(9.7)}
$$
Clearly we get $B_0=0$, while
$$
\dot{B}_t=e^{-iKt}C_te^{iKt}\ ,\ C_t=\dot{A}_t^\i-i[K,A_t^\i]\ .
\lo{(9.8)}
$$
Here the expression $C_t$ belongs to $Op\psi q_{-\i}$,
since its symbol satisfies (9.2) modulo $\psi q_{-\i}$.
It follows that
$$
e^{-iKt}A_t^\i e^{iKt}-A_0^\i=B_t=\int_0^t d\t e^{-iK\t}C_\t e^{iK\t}\ ,\
\lo{(9.9)}
$$
hence
$$
e^{iKt}A_0^\i e^{-iKt}=A_t^\i -\int_0^t e^{i(t-\t)K}C_\t e^{-i(t-\t)K}\ .
\lo{(9.10)}
$$
Here the operators $e^{\pm i(t-\t)K}$ are of order 0, in the sense of sec.3, while $C_t\in Op\psi q_{-\i}$ is of order $-\i$,
also in the sense of sec.3. Thus the term $-\int_0^t e^{i(t-\t)K}C_\t e^{-i(t-\t)K}$ also is of order $-\i$, while the 
$Op\psi q_{-\i}$-asymptotic limit of $A_t^\i$ may be written as an asymptotic sum, 
in the sense of sec.3 . Therefore we have obtained the desired asymptotic
expansion  of thm.4.1 with explicit operators $A_j$ given from the $Op\psi q_{-\i}$-asymptotic limit of $A_t^\i$.

We shall discuss our iteration in details in sec.10, below.

\LARGE

\section{Details Regarding the Iteration}

\normalsize

Let us discuss the iteration --- applying the 3 steps of sec.9.

\begin{lemma}
We have
$$
p_\pm(\xi)\a_jp_\pm(\xi)=\pm s_j(\xi)p_+(\xi)\ ,\
p_\pm(\xi)\b p_\pm(\xi)=\pm s_0(\xi)p_+(\xi)\ ,\
\lo{(10.1)}
$$
where we have set $s_j(\xi)=\xi_j/\<\xi\>\ ,\ j=1,2,3\ ,\ s_0(\xi)=1/\<\xi\>$ \ .\ 

\end{lemma}

The proof is a calculation, using the special properties of our Dirac matrices $\a_j,\b$ : $\a_j\a_l+\a_l\a_j=2\d_{jl}\ ,\
\a_j\b+\b\a_j=0\ .$

\begin{prop}

The operation $c(x,\xi)\rta (Xc)(x,\xi)$ (with $X$ of (9.3))
lowers the differentiation
order $m$ of $c\in \psi q_m$ by one unit -- to $\psi q_{m-1}$.

Also, if a symbol $M(x,\xi)$ commutes with $h_0(\xi)=\a\xi+\b$ then
we get

$(p_+[\a_2,M]p_+)(x,\xi)=(p_-[\a_2,M]p_-(x,\xi))=0\ .$

\end{prop}

We then start with the observables $D_j$, setting $q(x,\xi)=\xi_j$  --- of order $1$, commuting with $h_0(\xi)$.
We shall have to add a lower order $z(x_1,\xi)$ (of order $0$) , setting $a(x_1,\xi)=q(\xi) + z(x_1,\xi)$ , 
to get our approach working.

Looking at (9.2), seeking to omit all terms of lower order, and assuming $a_t(x_1,\xi)=q_t(x_1,\xi)+z_t(x_1,\xi)$ , as proposed
(with $[h_0,q_t]=0\ ,\ z_t$ of lower order), we get the simplified equation

$$
\dot{q}_t=i[h_0,z_t]+(\a_1-1)q_{t|x_1}+Z(q_t)\ \ (\mbox{  mod }\psi p_0 )\ .
\lo{(10.2)}
$$

Here we apply the multiplication $p_+\{XX\}p_+$ of `step II', noting that
$p_+[h_0,z_t]p_+=0$, and that
$p_+Z(q_t)p_+\in\psi p_0$, due to prop.10.2, also using lemma 10.1,  so that (10.2) simplifies to
$$
\dot{q}^+_t=(s_1-1)q^+_{t|x_1}\ \ (\mbox{  mod }\psi p_0)\ .
\lo{(10.2')}
$$
The sharp D.E. (10.2') with initial-value $q^+_0(x_1.\xi)=\xi_jp_+(\xi)$
has the unique solution $q_t^+(x_1,\xi)=\xi_jp_+(\xi)$.
Similarly we get $q_t^-(x_1,\xi)=\xi_j p_-(\xi)\ .$

So, we will get just
$$
q_t(x_1,\xi)=q_t^+(\xi)+q_t^-(\xi)=\xi_j(p_+(\xi)+p_-(\xi))=\xi_j
\ ,\ j=1,2,3\ .
\lo{(10.3)}
$$
Next we apply step III - multiplying $p_+\{XX\}p_-$ with
$a_t=q(\xi)+z_t(x_1,\xi)$ in (9.2), using that $q_t=\xi_j$ is a scalar independent of
$x$ and $t$, and that
$$
p_+[h_0,c]p_-=2\<\xi\>c^\pm\ ,\ p_-[h_0,c]p_+=-2\<\xi\>c^\mp\ ,
\lo{(10.4)}
$$
we get
$$
\dot{z}_t^\pm=2i\<\xi\>z_t^\pm +((\a_1-1)z_{t|x_1})^\pm
-2i\e_0s_2(\xi)\sin\o x_1\ z_t^\pm\}
+\frac{\e_0}{2}(\a_2Xa_t)^\pm\ .
\lo{(10.5)}
$$
Assuming that $\dot{z}_t$ also is of order $0$
and omitting all terms of order $0$ this reads
$$
2i\<\xi\>z_t^\pm =0\ \ \ \mbox{  (modulo }\psi q_0)\ .\ 
\lo{(10.5')}
$$
Since division by $\<\xi\>$ lowers the order by $1$
we thus get (also, repeating the procedure with $p_-\{XX\}p_+$)
$$
z_t^\pm=z_t^\mp=0 \mbox{  (mod }\psi q_{-1} )\ .
\lo{(10.6)}
$$
Both, $z_t^\pm$ and $z_t^\mp$ are approximations modulo $\psi q_{-1}$,\ ,
to be improved in the next iteration.

\begin{rem}

Note that our above condition of $\dot{z}_t\in\psi q_0$ is satisfied by our choice (10.6) of $z_t^\pm$, 
so that the construction is in order. 

\end{rem}

\bigskip

For the next iteration we return to steps I and II: With above $q_t=\xi_j$ and $z_t=0$, setting
$$
a_t=q +w_t+v_t \ ,\ \mbox{ where } w_t\in\psi q_0\ ,\ v_t\in \psi q_{-1}\ ,\ w_t=w_t^++w_t^-\ ,\ v_t=v_t^\pm+v_t^\mp\ .
\lo{(10.7)}
$$
Substituting into (9.2), using that $\dot{q}=q_{|x_1}=[h_0,q+w_t]=0$, we get
$$
\dot{a}_t=\dot{w}_t+\dot{v}_t=i[h_0,v_t]+(\a_1-1)(w_{t|x_1}+v_{t|x_1})+Z(q+w_t+v_t)\ . 
\lo{(10.8)}
$$
Assuming again $\dot{v}_t$ of order $-1$, multiplying $p_+\{XX\}p_+$, and omitting terms of order $-1$, we get
$$
\dot{w}_t^+=(s_1(\xi)-1)w_{t|x_1}^+ +\a_1^\pm(\xi)v^\mp_{t|x_1}+(Z(q+w_t+v_t))^+(x,\xi)\ ,\ \mbox{  (mod } \psi q_{-1} )
\lo{(10.9)}
$$
where we used lemma 10.1, and that $p_+\a_1cp_+=\a_1^+c^++\a_1^\pm c^\mp$ .

We still simplify 
$(Zc)(x_1,\xi)=-i\e_0\sin\o x_1 [\a_2,c(x_1,\xi)]
+\frac{\e_0}{2}\a_2(Xc)(x_1,\xi)\ ,\ c=\xi_j+w_t+v_t$, noting that $[a_2,\xi_j+w_t]^+=0$ and that $X(w_t+v_t)$ is of order $-1$,
by prop.10.2. So, we get
$$
\dot{w}_t^+=(s_1(\xi)-1)w_{t|x_1}^+ +\frac{\e_0}{2}(\a_2X(\xi_j))^+(x_1,\xi)\ ,\ \mbox{  (mod } \psi q_{-1} )\ .
\lo{(10.10)}
$$

\bigskip

Relation (10.10) again will be regarded as a sharp differential equation for
$w_t^+$. We may write it as
$$
\p_tw^+_t(x_1-t(s_1(\xi)-1),\xi)=F_t(x_1-t(s_1(\xi)-1),\xi)\ ,\
\lo{(10.11)}
$$
$$    
\mbox{ with   }
F_t(x_1,\xi)=\frac{\e_0}{2}(\a_2X(\xi_j))^+(x_1,\xi)=0\ ,\ \mbox{  as }j=2,3\ ,\ =\o\e_0 s_2(\xi)p_+(\xi)\cos(\o x_1)
\ ,\ \mbox{  as  } j=1\ .
$$
This (with initial value $w^+_0(x_1,\xi)$) is solved by integration; we get
$$
w_t^+(x_1-t(s_1(\xi)-1),\xi)=w_0^+(x_1,\xi)
+\int_0^t d\t F_\t(x_1-\t(s_1(\xi)-1),\xi)\ .\
\lo{(10.12)}
$$

Substituting $x_1-t(s_1(\xi)-1)$ by $x_1$ will give us
$$
w^+_t(x,\xi)=w^+_0(x_1+t(s_1(\xi)-1),\xi)
+\int_0^t d\t F_\t(x_1+(t-\t)(s_1(\xi)-1),\xi)\ .
\lo{(10.13)}
$$
We assume $w^+_0=0$ as to leave the original commutative
part $q=q_0$ untouched. Then we get
$$
w^+_t(x_1,\xi)=\int_0^t d\t F_{t-\t}(x_1+\t(s_1(\xi)-1),\xi)\ .
\lo{(10.14)}
$$
So, for the momentum components $D_2\ ,\ D_3$ we again get $w_t^+=0$. For $D_1$ we get
$$
w_t^+(x_1,\xi)= \o\e_0 s_2(\xi)p_+(\xi)\int_0^t d\t\cos(\o ((x_1-\t)+\t s_1(\xi)))\ .\ 
\lo{(10.15)}
$$
Clearly, for the product $p_-\{XX\}p_-$ the same derivation (from (10.8) to (10.15)) will apply, but we must replace 
$s_j(\xi)$ by $-s_j(\xi)$, according to lemma 10.1.
We get
$$
w_t^-(x_1,\xi)= -\o\e_0 s_2(\xi)p_-(\xi)\int_0^t d\t\cos(\o ((x_1-\t)-\t s_1(\xi)))\ .\ 
\lo{(10.15-)}
$$

Having obtained our $w=w^+ + w^-$ we return to (10.8) with the multiplications $p_+\{XX\}p_-$ etc. of step III.
Here the first term at right $ip_+[h_0,v_t]p_-=2i\<\xi\>v^\pm$ still will be of order $0$; we may ignore all terms of order $-1$ :
$$
\dot{v}_t^\pm=2i\<\xi\>v_t^\pm+\a_1^\pm w_{t|x_1}^-+Z(q+w_t)^\pm \mbox{  (mod }\psi q_{-1} )\ .\ 
\lo{(10.16)}
$$
With our assumption that also $\dot{v}_t\in\psi q_{-1}$ we then get
$$
v_t^\pm=-\inv{2i\<\xi\>}\{\a_1^\pm w_{t|x_1}+Z(q+w_t)^\pm\}\ \ \ \ ,\ \ \ \ 
v_t^\mp=\inv{2i\<\xi\>}\{\a_1^\mp w_{t|x_1}+Z(q+w_t)^\mp\}\ ,\ 
\lo{(10.17)}
$$
 In particular, our conditions on $\dot{w}_t\ ,\ \dot{v}_t$ are satisfied,so that
our construction is meaningful.

It can be seen now, how this iteration works: Writing $a_t^1(x_1,\xi)$ for our present $q+w_t+v_t$ we introduce
$a^2_t=a^1_t+W_t+V_t$ with $W_t=W_t^+ + W_t^-\in \psi q_{-1}\ ,\ V_t=V_t^\pm+V_t^\mp\in \psi q_{-2}$, 
assuming $\dot{W}_t\ ,\ \dot{V}_t$
of the same order than $W_t,V_t$, resp.. Substituting into (9.2), ignoring terms of lower order, applying our multiplications
of step II and step III we first obtain a differential equation in the variables $t,x_1$ --- always in the form (10) , solvable
in the form (11), for the $W_t^+\ ,\ W_t^-$, allowing us to determine $W_-$, then a commutator equation for the $V_t$, allowing 
to construct $V_t$ of the proper order, hence a new approximation of order $-1$, etc.

In this way we obtain an asymptotic expansion modulo $Op\psi q_{-\i}$, also implying the same expansion modulo $\mclh_{-\i}$.
Since we have $U(t)=T_{-t}e^{-iKt}$ and $T_tD_jT_{-t}=D_j$ we get $U^{-1}(t)(T_{-t}AT_t)U(t)$ as the desired expanasion 
for the Heisenberg transform: with pp-correction given by $z(x_1-t,D)$. The operator $U^{-1}(t)z(x_1-t)U(t)$ then will be of 
$\mclh_s$-order $-1$ to be integrated with the term $A^2_{jt}$. 
This should be a complete argument for proving thm.4.1.

\bigskip

\LARGE

\begin{center}

References   

\end{center}

\medskip

\footnotesize

\noindent
[Be1] R. Becker, \emph{Theorie der Electrizitaet}; Bd.2, B.G. Teubner
Verlag, Leibzig 1949.

\noindent
[BLT] N. N. Bogoliubov, A. A. Logunov and I. T. Todorov,
{\em Introduction to Axiomatic Quantum Field Theory}, Benjamin, 

Reading, Massachusetts, 1975.

\noindent
[Bu1] V.S.Buslaev, \emph{The generating integral and the canonical Maslov operator
in the WKB-method}; Funct. anal. iego pril., 3:3 (1969), 17-31. English translation: Funct. Anal. Appl., 3 (1969), 181-193.

\noindent
[CZ] A.P. Calderon and A.Zygmund, \emph{Singular integral operators and
differential equations}; Amer. J. Math. 79 (1957) 

 901-921.

\noindent
[Cp1] A.Compton, Phys.Rev. 21 483 (1923).

\noindent
[Cp2] A.Compton, Phil.Mag. 46 897 (1923).

\noindent
[Co1] H.O.Cordes, A pseudo-algebra of observables for the Dirac equation;
Manuscripta Math. 45 (1983) 77-105.

\noindent
[Co2] H.O.Cordes, {\em The technique of pseudodifferential operators};
London Math. Soc. Lecture Notes 202; Cambridge Univ. 

Press 1995, Cambridge.

\noindent
[Co3] H.O.Cordes, {\em Elliptic pseudo-differential operators - an abstract
theory}; Springer Lecture Notes Math. Vol. 756, 

Springer Berlin Heidelberg New York 1979

\noindent
[Co4] H.O.Cordes, {\em Spectral theory of linear differential operators
and comparison algebras}; London Math. Soc. Lecture 

Notes No.76 (1987); Cambridge Univ. Press; Cambridge.

\noindent
[Co5] H.O.Cordes, \emph{The split of the Dirac Hamiltonian into precisely
predictable energy components}; Fdns. of Phys. 34 

(1004) 1117-1153.

\noindent
[Co6] H.O.Cordes, \emph{Precisely predictable Dirac Observables}; Fundamental
Theories of Physics 154 Springer 2007.

\noindent
[Co7] H.O.Cordes, \emph{A mathematical analysis of Dirac equation Physics}; Investigations
Math. Sci., 4(2) 2014 1-53.

\noindent
[DEFJKM] P.Deligne, P.Etingof, D.Freed, L.Jeffrey, D.Kazhdan, and
D.Morrison, {\em Quantum fields and Strings for 

Mathematicians}; Princeton Univ. Press; Princeton 1999.

\noindent
[FS] L.D.Faddeev and A.A.Slawnov, {\em Gauge fields}; Introduction to
Quantum Theory; 

Benjamin/Cummings 1980 Reading MA London Amsterdam Sydney Tokyo.

\noindent
[FW] L. Foldy, S. Wouthuysen, \emph{On the Dirac theory of spin $-\halb$
particles}.  Phys Rev 78:20-36, 1950.

\noindent
[GS] I. Gelfand and G.E.Silov, {\em Generalized Functions},
Vol.1; Acad. Press New York 1964.

\noindent
[GL] M.Gell-Mann and F.Low, \emph{Quantum electrodynamics at small distances};
 Phys. Rev. 95 (1954) 1300-1312 .

\noindent
[Go1] I. Gohberg, \emph{On the theory of multidimensional singular integral
operators}; Soviet Math. 1 (1960) 960-963.

Basel 1979 (Russian ed. 1973).

\noindent
[Hd1] J. Hadamard, {\em Lectures on Cauchy's problem}; Dover, New York 1953
[Originally published by Yale Univ.Press in 

1923].

\noindent
[Hi1] D. Hilbert, {\em Integralgleichungen}; Chelsea NewYork 1953.

\noindent
[HLP] G.H.Hardy, J.E.Littlewood, and G.Polya, {\em Inequalities}; Cambridge
Univ. Press 1934.

\noindent
[Hs1] W. Heisenberg.  {\em Gesammelte Werke}.
Berlin-New York: Springer, 1984.
    
\noindent
[Hoe1] L. Hoermander, {\em Linear partial differential operators};
Springer New York Berlin Heidelberg 1963.

\noindent
[Hoe2] L.Hoermander, \emph{Pseudodifferential operators and hypo-elliptic
equations}; Proceedings Symposia pure appl. Math. 10 

(1966) 138-183.

\noindent
[Hoe3] L.Hoermander, {\em The analysis of linear partial
differential operators} Vol's I--IV; Springer New York Berlin Heidelberg 

1983-1985.

\noindent
[Hoe4] L.Hoermander, {\em Fourier integral operators I}; Acta.math. 127 (1971) 79-183.

\noindent
[Ka1] T.Kato, {\em Perturbation theory for linear operators};
Springer Verlag Berlin Heidelberg New York 1966.

\noindent
[LS] Laurent Schwartz, {\em Theorie des distributions}; Herman Paris 1966.

\noindent
[MO] W.Magnus and F.Oberhettinger, {\em Formeln und Saetze fuer die
speziellen Funktionen der Mathematischen Physik}; 

2.Auflage, Springer Verlag Berlin Goettingen Heidelberg 1948.

\noindent
[MOS] W.Magnus, F.Oberhettinger and R.P.Soni, {\em Formulas and theorems
for the special functions of Mathematical Physics}; 

3rd edition, Springer Verlag New York 1966.

\noindent
[Ms1] V.P.Maslov, \emph{Theory of perturbations and asymptotic methods};
Moskow Gos. Univ. Moskow, 1965.

\noindent
[Mu] C. M\"uller, {\em Grundprobleme der Mathematischen Theorie
elektromagnetischer Schwingungen}; Springer Verlag, Berlin 

G\"ottingen  Heidelberg 1957.

\noindent
[JvN] J.v.Neumann, {\em Die Mathematischen Grundlagen der Quantenmechanik};
 Springer 1932 New York; reprinted Dover. Publ. inc. 1943; English
translation 1955 Princeton Univ. Press.

\noindent
[Sa] A.Salam, {\em Elementary Particle Theory} N.Svartholm (ed)
Stockholm Almquist Forlag AB 1968

\noindent
[Schr\"{o}1] E. Schr\"{o}dinger.\emph{\"{U}ber den Comptoneffekt;}
Annalen der Physik (4) 82 (1927)

\noindent
[Schr\"{o}2] E. Schr\"{o}dinger.\emph{Quantisierung als
Eigenwertproblem;} Annalen der Physik (4) 79 (1926)

\noindent
[Schr\"{o}3] E. Schr\"{o}dinger.\emph{Collected Papers;}
Friedr. Viehweg und Sohn 1084.

\noindent
[SB] D.Shirkov, N.Bogoliubov. {\em Quantum Fields}.
Reading, MA: Benjamin, 1982.

\noindent
[So1] A.Sommerfeld, {\em Atombau und Spektrallinien},
vol.1. 5th ed. Braunschweig, Viehweg and Sons, 1931.

\noindent
[So2] A.Sommerfeld, \emph{Atombau und Spektrallinien}, Vol.2.
Braunschweig Vieweg and Sons, 1931.

\noindent
[Ta1] M. Taylor, {\em Pseudodifferential operators}; Princeton Univ. Press.,
Princeton, NJ 1981.

\noindent
[Ta2] M.Taylor, {\em Partial differential equations}; Vol.I,II,III;
Springer New York Berlin Heidelberg 1991.

\noindent
[Th1] B.Thaller, {\em The Dirac equation}; Springer 1992 Berlin Heidelberg
New York.

\noindent
[Ti1] E.C.Titchmarsh, {\em Eigenfunction expansions associated with
second order differential equations} Part 1, 2-nd ed.

Clarendon Press, Oxford 1962.

\noindent
[Ti2] E.C.Titchmarsh, {\em Eigenfunction expansions associated
with second order differential equations} Part 2 [PDE];
Oxford 

Univ. Press 1958.

\noindent
[Un1] A. Unterberger, \emph{A calculus of observables on a
Dirac particle},
Annales Inst. Henri Poincar\'e (Phys. Th\'eor.), 69 (1998) 

189-239.

\noindent
[Un2] A. Unterberger, \emph{Quantization, symmetries and
relativity}; Contemporary Math. {\bf 214}, AMS (1998), 169-187.

\noindent
[Wa1] G.N.Watson, \emph{A Treatise on the Theory of Bessel Functions};
Cambridge Univ. Press, 1922.

\noindent
[Wb1] S.Weinberg, \emph{on weak forces and gauge theory with SU(2)};
Phys. Rev. Lett. 19 (1967) 1264.

\noindent
[We1] A.Weinstein, \emph{A symbol class for some Schr\"odinger equations on
$\mbb{R}^n$}; Amer. J. Math. (1985) 1-21.

\noindent
[Wi1] E. Wichmann. {\em Quantenphysik}.  Braunschweig: Viehweg und Sohn, 1985.

\noindent
[YM] C.N.Yang and R.L.Mills, \emph{Conservation of isotopic spin and isotopic
gauge invariance}; Phys.Rev. 96 (1954) 191-195.

\normalsize
\vspace {1 cm}
\noindent
Emeritus Professor\\
Department of Mathematics\\
University of California\\
Berkeley, CA 94720, U.S.A.\\
E-mail: cordes@math.berkeley.edu\\

\end{document}